\documentclass[twocolumn,10pt,prl,aps,showpacs,superscriptaddress,preprintnumbers,amsmath,amssymb]{revtex4-1}
\usepackage{graphicx}
\usepackage{dcolumn}
\usepackage{bm}
\usepackage{eqnarray}
\usepackage{color}
\usepackage{bigstrut}
\usepackage{tabularx}
\usepackage{float}
\usepackage[]{color}
\usepackage{silence}
\usepackage{graphicx,amssymb}
\usepackage{color}
\usepackage[T1]{fontenc}
\usepackage{xcolor}
\usepackage[colorlinks,citecolor=red,urlcolor=blue,bookmarks=false,hypertexnames=true]{hyperref} 
\usepackage[section]{placeins}
\usepackage{gensymb}
\usepackage{amsmath}
\usepackage{multirow}
\usepackage[utf8]{inputenc} 
\usepackage[T1]{fontenc}

\makeatletter
\renewcommand{\section}{%
  \@startsection{section}{1}{0 pt}%
    {-1ex plus -.5ex minus -.2ex} 
    {0pt} 
    {\large\bfseries}%
}

\usepackage{titlesec}

\titleformat{\section}[hang]
  {\normalfont\large\bfseries} 
  {} 
  {0pt} 
  {} 

\titleformat{\subsection}[runin] 
  {\normalfont\normalsize\bfseries} 
  {} 
  {0.5em} 
  {} 

\titlespacing{\subsection}
  {0pt} 
  {1ex} 
  {0pt} 

\makeatother

\begin{document}
\bibliographystyle{apsrev}
\newcommand{\red}[1]{\textcolor{red}{#1}}

\WarningFilter{revtex4-1}{Repair the float}

\setlength{\fboxrule}{0.5pt}

\title{Local electronic properties of $\rm La_3Ni_2O_7$ under pressure}

\author{Emin Mijit}
\email{emin.mijit.17@gmail.com}
\affiliation{ESRF, The European Synchrotron\unskip, 71 Avenue des Martyrs\unskip, CS40220\unskip, 38043 Grenoble Cedex 9\unskip, France}

\author{Peiyue Ma}
\affiliation{Center of Neutron Science and Technology\unskip, Guangdong Provincial Key Laboratory of Magnetoelectric Physics and Devices\unskip, School of Physics\unskip, Sun Yat-Sen University\unskip, Guangzhou\unskip \ 510275, China}

\author{Christoph J. Sahle}
   \affiliation{ESRF, The European Synchrotron\unskip, 71 Avenue des Martyrs\unskip, CS40220\unskip, 38043 Grenoble Cedex 9\unskip, France}

\author{Angelika D.Rosa}
\email{angelika.rosa@esrf.fr}
\affiliation{ESRF, The European Synchrotron\unskip, 71 Avenue des Martyrs\unskip, CS40220\unskip, 38043 Grenoble Cedex 9\unskip, France}

\author{Zhiwei Hu}
   \affiliation{Max-Planck Institute for Chemical Physics of Solids - Nothnitzer Straße 40\unskip, Dresden 01187\unskip, Germany}

\author{Francesco De Angelis}
   \affiliation{Department of Science Roma Tre University\unskip, Via della Vasca Navale 79\unskip, 00146\unskip, Rome\unskip, Italy}
   
\author{Alberto Lopez}
   \affiliation{Department of Science Roma Tre University\unskip, Via della Vasca Navale 79\unskip, 00146\unskip, Rome\unskip, Italy}

\author{Simone Amatori}
   \affiliation{Department of Science Roma Tre University\unskip, Via della Vasca Navale 79\unskip, 00146\unskip, Rome\unskip, Italy}

\author{Georghii Tchoudinov}
\address{Physics Division, School of Science and Technology, Universit\`a di Camerino, Via Madonna delle Carceri 9, I-62032 Camerino (MC), Italy} 

\author{Yves Joly}
\address{Universit\'e Grenoble Alpes, CNRS, Institut N\'eel, F-38042 Grenoble, France} 

\author{Tetsuo Irifune}
\affiliation{Geodynamics Research Center\unskip, Ehime University\unskip, Matsuyama 790\--8577\unskip, Japan}

 \author{Jo$\tilde{a}$o Elias F. S. Rodrigues}  
\affiliation{ESRF, The European Synchrotron\unskip, 71 Avenue des Martyrs\unskip, CS40220\unskip, 38043 Grenoble Cedex 9\unskip, France}

\author{Gaston Garbarino}
\affiliation{ESRF, The European Synchrotron\unskip, 71 Avenue des Martyrs\unskip, CS40220\unskip, 38043 Grenoble Cedex 9\unskip, France}

\author{Samuel Gallego Parra}
\affiliation{ESRF, The European Synchrotron\unskip, 71 Avenue des Martyrs\unskip, CS40220\unskip, 38043 Grenoble Cedex 9\unskip, France}

\author{Meng Wang}
\email{wangmeng5@mail.sysu.edu.cn}
\affiliation{Center of Neutron Science and Technology\unskip, Guangdong Provincial Key Laboratory of Magnetoelectric Physics and Devices\unskip, School of Physics\unskip, Sun Yat-Sen University\unskip, Guangzhou\unskip \ 510275, China}

\author{Runze Yu}
\email{runze.yu@hpstar.ac.cn}
  \affiliation{Center for High Pressure Science $\&$ Technology Advanced Research - Beijing 100094\unskip, China}

\author{Olivier Mathon}
\email{mathon@esrf.fr}
\affiliation{ESRF, The European Synchrotron\unskip, 71 Avenue des Martyrs\unskip, CS40220\unskip, 38043 Grenoble Cedex 9\unskip, France}

\date{\today}

\begin{abstract}
The recent discovery of superconductivity in $\rm La_3Ni_2O_7$ has attracted significant attention due to its high critical temperature and analogy to cuprate oxides. The oxidation and spin states of Ni ions are among the most important local properties in this compound, extensively discussed in the context of its superconductivity. Despite their direct link to the electron filling configurations of the relevant $\rm 3d_{x^2-y^2}$ and $\rm 3d_{z^2}$ orbitals, these local electronic properties of $\rm La_3Ni_2O_7$ yet to be systematically investigated. In this work, we address this issue using x-ray absorption spectroscopy (XAS) and x-ray emission spectroscopy (XES) measurements under pressure. Comparison of Ni \textit{K}-edge XAS and $\rm K\beta$ XES with the reference spectra of $\rm NiO$ and $\rm LaNiO_3$ shows that Ni ions, with an average valence of $\sim 2.53+$, are in a low-spin ($\rm S = 1/2$) ground state under ambient conditions. High pressure XAS and XES data clearly show that the oxidation ($\sim 2.5+$) and spin ($\rm S = 1/2$) states of Ni ions remain stable across the investigated pressure (up to 30 GPa) and temperature (down to 10 K) ranges, ruling out previously proposed spin transition scenarios. 
\end{abstract}

\maketitle

\section{Introduction}
\ Superconductivity in cuprate oxides is known to be associated with the Cu-O layers, where $\rm Cu^{2+}$ ions possess a $\rm 3d^9$ electronic configuration and a spin state of $\rm S=1/2$ \cite{Patrick2006}, leaving an unpaired electron and a single hole in the 3d shell for each unit cell. Due to their structural and electronic similarities to cuprates, infinite-layer nickelates with $\rm Ni^{1+}$ cations ($\rm 3d^9$) have been extensively investigated as candidates for high transition temperature ($\rm T_c$) superconductors \cite{Li2019,Osada2020,Pan2022,Wang2022}. The highest $\rm T_c$ achieved with this family of compounds is at 31 K, observed in $\rm Pr_{0.82}Sr_{0.18}NiO_2$ thin films under pressure ($\sim$ 12.1 GPa) \cite{Wang2022}, which is still below the so-called McMillan limit of 40 K(the theoretical maximum $\rm T_c$ for conventional superconductivity \cite{McMillan1968}). 

Surprisingly, superconductivity exceeding 80 K has recently been observed in $\rm La_3Ni_2O_7$ \cite{Sun2023}, a layered nickelate with an average electron occupancy of $\rm 3d^{7.5}$ ($\rm Ni^{2.5+}$). This marks the discovery of a second system (after cuprates) that shows superconducting transitions above the boiling temperature of nitrogen under ambient or moderate high pressures. For a Ni ion with $\rm 3d^{7.5}$ occupancy in octahedral coordination, six electrons fully occupy three $\rm t_{2g}$ orbitals, leaving an average of 1.5 electrons for the two $\rm e_{g}$ orbitals. At first glance, this electronic arrangement seems incompatible with the cuprate-like orbital configurations. However, density functional theory (DFT) calculations have shown that strong interlayer coupling further splits two $\rm e_{g}$ ($\rm 3d_{x^2-y^2}$ and $\rm 3d_{z^2}$) orbitals \cite{Pardo2011,Nakata2017,Sun2023,Luo2023bilayer}. Based on nonmagnetic DFT calculations, a bilayer model with half filled $\rm 3d_{z^2}$ and quarter-filled $\rm 3d_{x^2-y^2}$ orbitals was suggested\cite{Sun2023}. Resembling the electronic structure of hole-doped bilayer cuprates \cite{SakakibaraPRB2014}, this filling configuration has been proposed as a plausible explanation for the superconductivity in $\rm La_3Ni_2O_7$ \cite{Sun2023}. Indeed, experimental studies using angle-resolved photoemission spectroscopy (ARPES) and XAS under ambient pressure have demonstrated the dominant roles of the $\rm 3d_{x^2-y^2}$ and $\rm 3d_{z^2}$ bands in the low-energy physics of the material \cite{Yang2024orbital,Chen2024electronic}. However, the above mentioned local electronic structure with doublet spin state ($\rm S=1/2$) for each bilayer (comprising two Ni ions), has yet to be experimentally verified. In fact, local spin state and electron filling configurations in those valence orbitals are still matter of discussion in many theoretical studies that investigate superconductivity in $\rm La_3Ni_2O_7$ \cite{Luo2023bilayer,Pardo2011,
Jiang2024pressure,Ouyang2024hund,Liao2023electron,Labollita2023electronic,ShilenkoPRB2023,Lin2024magnetic,Kakoi2024pair,Chen2024electronic,ChristianssonPRL2023, ChenOrbital2024,Chen2024non,Chen2023critical,LaBollitaPRL2024,Leonov2024electronic,Qin2024intertwined,ZhangPRB2023}. For example, based on DFT calculations including the local magnetic moment, $\rm S=3/4$ was obtained for $\rm 3d^{7.5}$ \cite{Pardo2011}. Hundness of electron correlations was suggested based on large percentage of high-spin (HS) ($\rm S=1$) configuration obtained by DFT+DMFT(dynamical mean field theory) calculations \cite{Ouyang2024hund}. Another study using band structure and DFT+DMFT calculations also obtained significant amount of HS state, yielding a formal valence of 1.75+ for Ni ions, being largely deviated from the expected value (2.5+) \cite{ShilenkoPRB2023}. Based on such quantities, this study has suggested that $\rm La_3Ni_2O_7$ is close to the negative charge transfer regime. Pressure induced formation of cuprate like electronic structure via fractionalizing Ni ionic spins (from $\rm S=1$ to $\rm S=1/2$) has been argued, suggesting d-wave symmetry for the pairing mechanism \cite{Jiang2024pressure}. Interestingly, opposite behavior, pressure induced low-spin (LS) to HS transition scenario was proposed by H. Labollia et al. based on correlated DFT (DFT+U) method \cite{Labollita2023electronic}.  Numerical simulations performed by Lioa et al. using bilayer Hubbard model have proposed that electron correlations drive the ground state from LS $\rm S = 1/2$ to HS $\rm S = 3/2$ one, indicating half-filled configuration for both $\rm 3d_{x^2-y^2}$ and $\rm 3d_{z^2}$ orbitals \cite{Liao2023electron}. Theoretical \cite{Chen2023critical,LaBollitaPRL2024,Qin2024intertwined,Leonov2024electronic} and experimental \cite{Chen2024electronic,Liu2023evidence, ChenPRL2024,Ren2024resolving,Xie2024strong} investigations studying density (charge or spin) wave states have discussed stripe models based on alternating $\rm Ni^{2+}$ and $\rm Ni^{3+}$ cations with different (LS/HS, or spin up/spin down/spinless) spin sates.

\begin{figure}[h]
\centering
\includegraphics[width=1.0\linewidth]{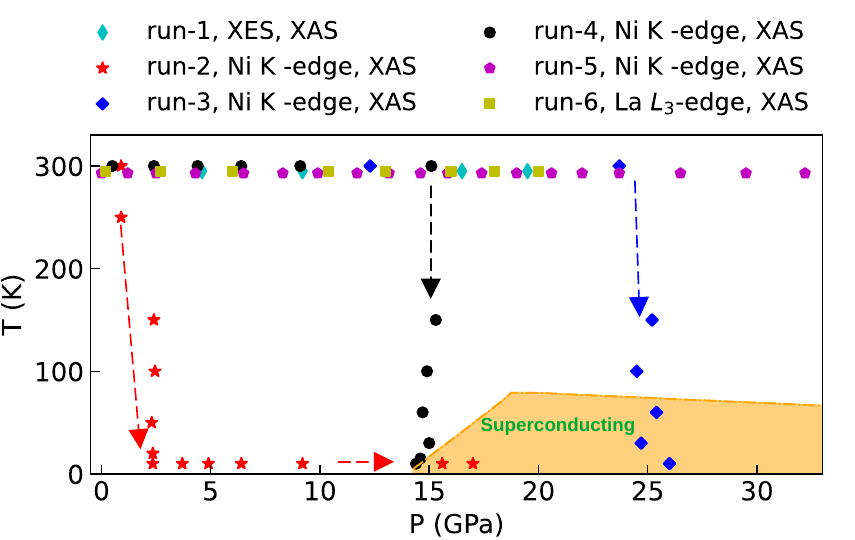}
\caption{Experimental conditions and routes mapped onto the phase diagram of $\rm La_3Ni_2O_7$ \cite{Sun2023, Wang2024structure}.}
\label{PD}
\end{figure}

Local spin and oxidation states are two fundamental measurable quantities that govern the electron filling configurations of the most relevant valence orbitals ($\rm 3d_{x^2-y^2}$ and $\rm 3d_{z^2}$ ) in $\rm La_3Ni_2O_7$. Based on above literature survey, it is evident that electronic properties of Ni ions in this compound require urgent investigation using microscopic techniques sensitive to local electronic structure. Given that pressure or temperature induced changes of oxidation and spin states are frequently observed in transition metal oxides \cite{Liu2020sequential,Leonov2016,Tkano1991,Pardo2012,Takegami2023}, it is essential to measure these properties in situ under high-pressure and low-temperature conditions, particularly within the superconducting domain (P > 14 GPa and T < 80 K). In this work, we address these pressing challenges by means of Ni \textit{K} and La \textit{$L_3$} -edges XAS and Ni $K_{\beta}$ XES measurements. Experimental conditions and routs that describe all pressure and temperature dependent measurements were mapped on the phase diagram in Fig. \ref{PD}.

\section {Results and discussion}

\subsection{At ambient conditions: }

\begin{figure}[!]
\centering
\includegraphics[width=0.95\linewidth]{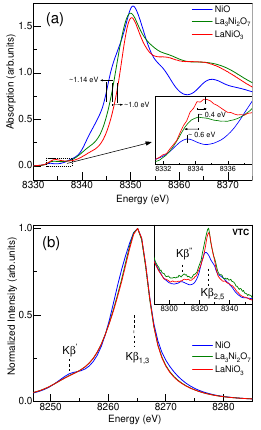}
\caption{Ni \textit{K}-edge XAS (a) and $K_{\beta}$ XES (b) spectra measured under ambient conditions. In (a), the energy positions of the absorption edge and pre-edge features (inset) are marked with vertical bars. The edge energy is defined as the point on the rising edge where the jump reaches 0.8 (normalized absorption). In (b), the inset shows satellite features in the valence-to-core (VTC) region. XES spectra were normalized by scaling the strongest $K\beta_{1,3}$ peak to 1.}
\label{Ref}
\end{figure}  

\begin{figure}[!]
\centering
\includegraphics[width=0.95\linewidth]{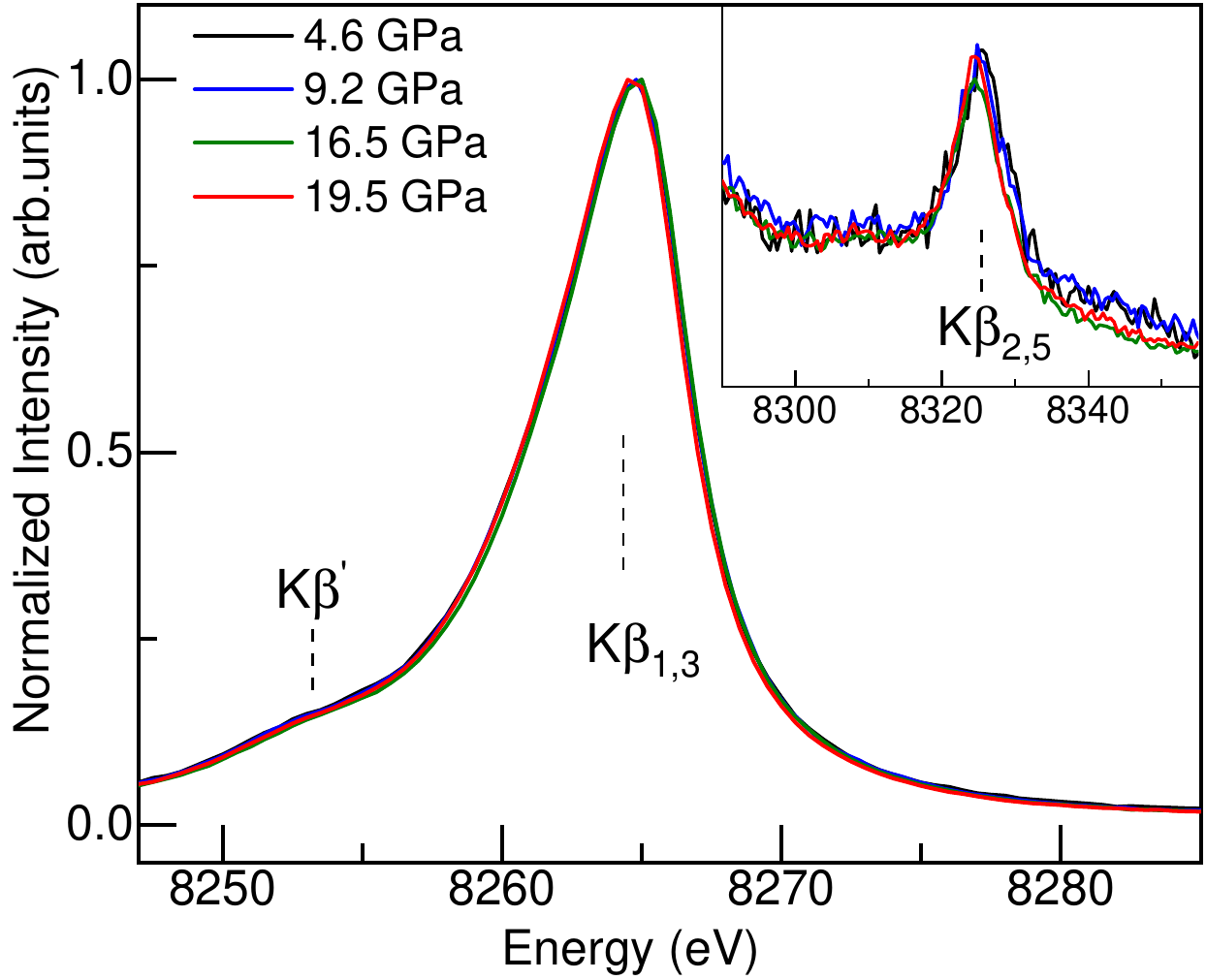}
\caption{Ni $K_{\beta}$ XES measured under pressure. All the data were normalized by scaling the strongest $K\beta_{1,3}$ peak to 1. } 
\label{XES}
\end{figure}

\begin{figure*}[t]
\centering
\includegraphics[width=0.95\linewidth]{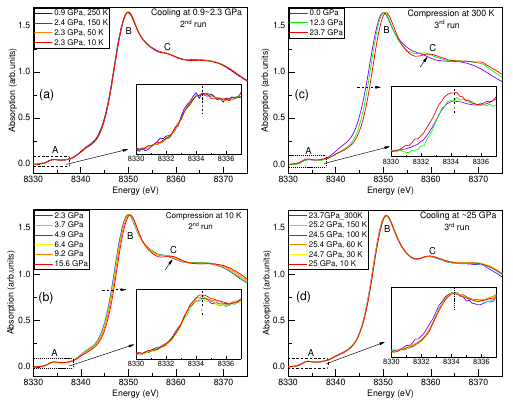}
\caption{Ni \textit{K} XAS measured under quasi-isobaric cooling (a,d) and isothermal compression (b,c) along two separate experimental runs (run-2 and run-3). Pre-edge ranges were plotted with magnification and given as a inset in each figure.} 
\label{Fig_2}
\end{figure*}

\ Ni \textit{K}-edge XANES (x-ray absorption near edge structure) and $K_{\beta}$ XES spectra of $\rm La_3Ni_2O_7$ measured at ambient conditions are compared with those of NiO and $\rm LaNiO_3$ reference compounds in Fig. \ref{Ref}. Both  NiO and $\rm LaNiO_3$ have octahedral local structure and representative of HS $\rm Ni^{2+}$ ($\rm S=1$) and LS $\rm Ni^{3+}$ ($\rm S=1/2$) ions, respectively. As shown in Fig. \ref{Ref}(a), XAS spectrum of $\rm La_3Ni_2O_7$ exhibit clear chemical shift relative to the spectra of NiO and $\rm LaNiO_3$. In particular, energy positions of the absorption edge and pre-edge feature are located in the middle of the corresponding features of NiO and $\rm LaNiO_3$ spectra. This indicates that the nickel oxidation state in $\rm La_3Ni_2O_7$ is intermediate between the $\rm Ni^{2+}$ in NiO and the $\rm Ni^{3+}$ in $\rm LaNiO_3$. Based on the chemical shifts observed at the rising edge, average valence of Ni is estimated to be around 2.53(5)+. This estimation assumes a linear relationship between the edge energy and oxidation state, and neglects possible energy shift coming from the difference in the spin state and local symmetry (distorted octahedra in case of $\rm La_3Ni_2O_7$).  

 From the $K_{\beta}$ XES spectra reported Fig. \ref{Ref}(b), we were surprised to observe that main emission lines ($K\beta_{1,3}$ and $K\beta^{\prime}$) of $\rm La_3Ni_2O_7$ overlap perfectly with those of $\rm LaNiO_3$ with LS  ($\rm S=1/2$) $\rm Ni^{3+}$. In contrast, $K\beta^{\prime}$ intensity is reduced with respect to NiO reference with HS ($\rm S=1$) spin state. No chemical shift was detected at the $K\beta_{1,3}$ main line, as its energy position is primarily governed by 3p-3d exchange splitting rather than nuclear screening effects \cite{Glatzel2005high}. However, in the valence to core emission region, the $K\beta_{2,5}$ peak (which is relatively more sensitivity to the oxidation state \cite{Bergmann1999chemical, Fazinic2011crossover}), has shown a noticeable chemical shift ($\rm \sim 1.50$ eV) compare to NiO. The shift relative to $\rm LaNiO_3$ was smaller ($\rm -0.35$ eV). Intensity and line shape of this satellite feature closely resemble those of $\rm LaNiO_3$, but differing clearly from NiO with reduced (25 \%) peak intensity. In 3d transition metal oxides, $K\beta_{2,5}$ line arises from dipole allowed transitions involving strong O-2p component and some metal 4p character \cite{Gallo2014valence}. Therefore, pronounced $K\beta_{2,5}$ peak in $\rm LaNiO_3$ reflects its negative charge transfer character, with leading $\rm 3d^{8}L$ (L: O 2p hole) contribution in the ground state \cite{Abbate2002,Chen2017charge,Takegami2024PRB}. This is because strong 2p mixing with Ni 3d states in a negative charge-transfer system may allow more p-like transitions for $K\beta_{2,5}$ emission. The nearly identical $K_{\beta}$ XES spectrum of $\rm La_3Ni_2O_7$ compared to $\rm LaNiO_3$ provide compelling evidence that $\rm La_3Ni_2O_7$ also exhibits LS ($\rm S=1/2$) ground state with substantial $\rm 3d^{8}L$ contribution. In addition, presence of HS configurations in $\rm La_3Ni_2O_7$ is incompatible with the present data. XES spectra with the characteristics of a single component LS system ($\rm LaNiO_3$) can not be explained with the presence of either pure or mixed HS state \cite{Ouyang2024hund, ShilenkoPRB2023,Chen2023critical,LaBollitaPRL2024,Qin2024intertwined,Leonov2024electronic,Liu2023evidence,ChenPRL2024}.

 \subsection{Under high pressure: }
 \ As one can see from Fig. \ref{XES}, no spectral modification was observed in the shape of the main $K\beta$ XES emission lines under pressure (run-1), indicating that the spin state remains stable up to 19.5 GPa. As shown in the inset of Fig. \ref{XES}, $K\beta_{2,5}$ satellite shifts to lower energies due to pressure effects \cite{Spiekermann2019,Spiekermann2023,Albers2023high}. However, its intensity and shape is almost the same at different pressure, suggesting that there are no significant changes in the amount of O 2p character mixed with Ni 3d orbitals. Information derived from the present XES data definitively rules out pressure-induced spin-state transition scenarios previously proposed in the literature \cite{Jiang2024pressure,Labollita2023electronic,Liao2023electron}. 
\begin{figure}[!]
\centering
\includegraphics[width=0.95\linewidth]{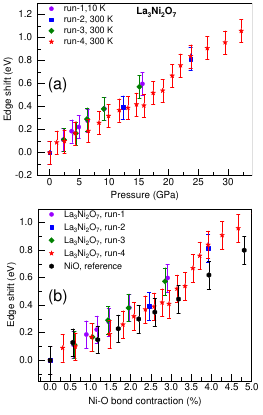}
\caption{Energy shift of Ni \textit{K}-edge as a function of pressure (a) and Ni-O bond contraction (b).}
\label{Edge_shift}
\end{figure}  

Ni \textit{K}-edge XANES spectra measured upon cooling down to 10 K ($\rm P\approx 0.9\sim 2.4$ GPa) during run-2, are presented in Fig. \ref{Fig_2}(a). No detectable spectral change was observed during this process, indicating the absence of significant local atomic or electronic modifications within this pressure-temperature (P-T) range. XANES spectra measured during subsequent isothermal compression at 10 K are shown in Fig. \ref{Fig_2}(b). As pressure increases up to 15.6 GPa, a gradual shift of the rising edge towards higher energies (approximately +0.5 eV at 15.6 GPa) is observed. Additionally, the feature around 8358 eV (labeled as \textbf{C}) becomes more pronounced, especially at the final pressure point of 15.6 GPa, when the system enters the superconducting phase. However, the pre-edge feature \textbf{A} shows no observable energy shift, but its intensity seem to be increased slightly under pressure [see the inset of Fig. \ref{Fig_2}(b)].

Similar behavior is observed in the third experimental run (run-3). As shown in Fig. \ref{Fig_2}(c,d), the XANES spectra exhibit a clear energy shift ($\sim 0.8$ eV at 23.7 GPa) at the rising edge and an enhancement of feature \textbf{C} upon isothermal compression at 300 K. Once again, pre-edge peak \textbf{A} remained stable at its initial energy position (within experimental uncertainty of $\pm 0.1$ eV), showing a slight increase in intensity. XANES profile did not change during quasi-isobaric cooling to 10 K around $23.7 \sim 25$ GPa [see Fig. \ref{Fig_2}(d)]. For a better insight into the spectral modifications, we plotted first derivative of all the XANES data collected along run-2. We refer Fig. S2 (SI file) and its caption for the details and interpretation of these first derivative spectra. Pressure behavior of the Ni \textit{K}-edge XANES in the forth and fifth experimental runs is consistent with what has been described above, so we omit detailed discussion of these data.

The energy shift of the absorption threshold ($\rm E_0$, referenced as the point on the rising edge where the absorption jump reaches 0.8) is plotted as a function of pressure in Fig. \ref{Edge_shift}(a), showing a slight anomaly at $13 \sim 15$ GPa. This anomaly likely corresponds to the orthorhombic (Amam) to tetragonal ($\rm I4/mmm$) structural phase transition around this pressures \cite{Wang2024structure,Li2024pressure}. 
The shift of $\rm E_0$ toward higher energy here is a delicate issue to be carefully interpreted, as it can arise from two different factors: (i) an increase in the average valence of Ni, or (ii) lattice contraction. To distinguish between these two possibilities, we conducted an additional compression experiment on a NiO reference for comparing the edge energy shift versus Ni–O bond distance contraction, as shown in Fig. \ref{Edge_shift}(b). Crystal and electronic structure of NiO with $\rm Ni^{2+}$ is known to be stable within this pressure range (up to 40 GPa) \cite{Potapkin2016}, thus its edge energy shift is of purely structural origin. As reported in Fig. \ref{Edge_shift}(b), the rate of energy shift versus Ni-O bond contraction in $\rm La_3Ni_2O_7$ is in line with one of NiO reference within the error bars. Therefore, possibility of noticeable change in the average Ni valence under high pressure can be excluded. 

\begin{figure}[!]
\centering
\includegraphics[width=0.9\linewidth]{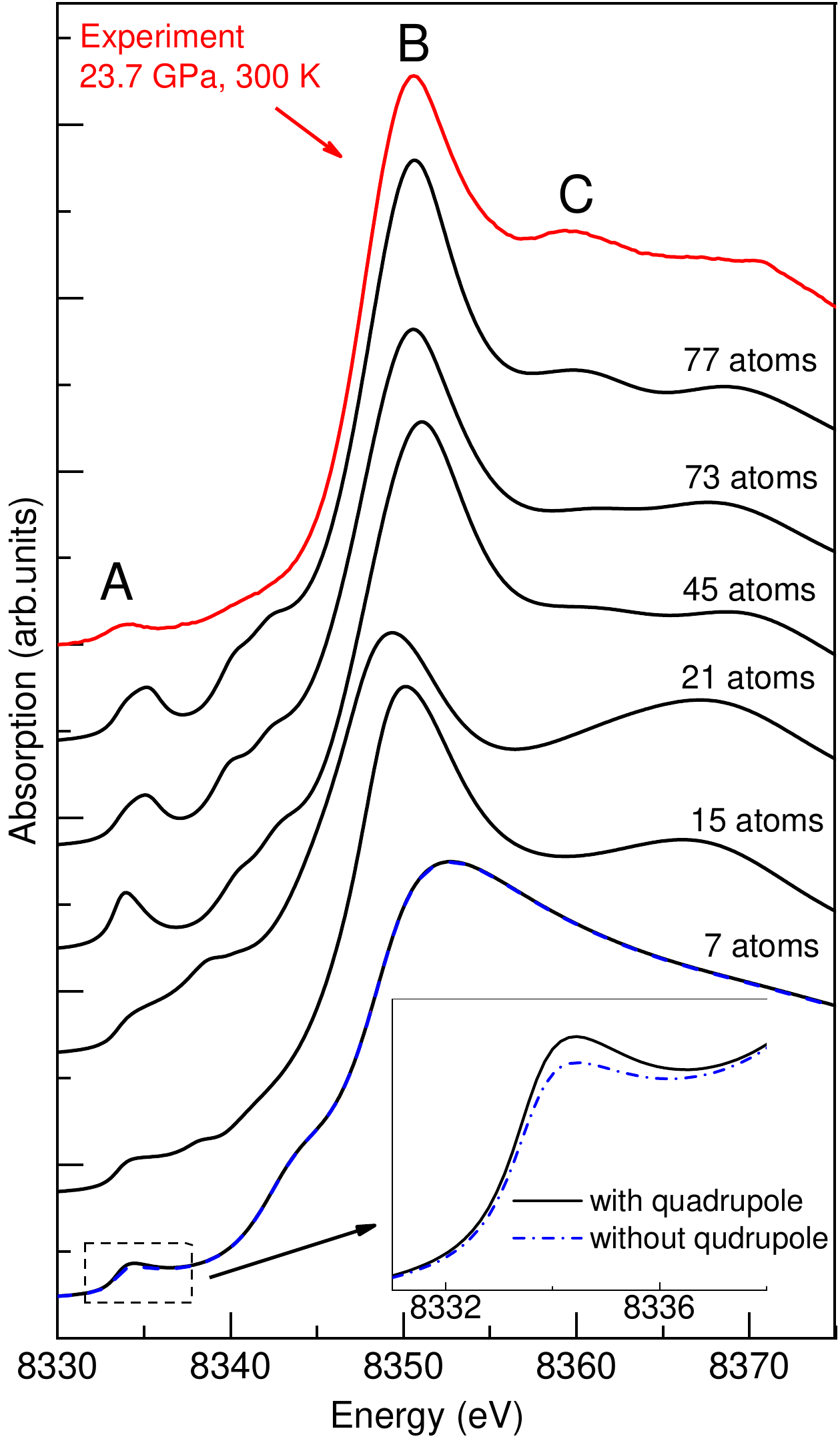}
\caption{XANES simulations performed for various atomic cluster sizes. Experimental spectra (red curve) was measured at P=23.7 GPa under room temperature.} 
\label{FDMNS}
\end{figure}

We performed XANES simulations using the FDMNES code \cite{FDMNS,Joly2001, Joly2009} to interpret spectral changes related to feature \textbf{A} and \textbf{C}. High-pressure $\rm I4/mmm$ phase \cite{Wang2024structure} was used to prepare input atomic clusters. Several simulations have been performed by gradually increasing the size of the spherical atomic cluster surrounding the absorber atom (Ni), adding successive atomic shells step by step. As shown in Fig. \ref{FDMNS}, pre-edge peak \textbf{A} was reproduced in the first calculation with a cluster of seven atoms, consisting solely of the $\rm NiO_6$ octahedron, showing its local origin. Pre-edge features in the \textit{K} edges of 3d transition metals arise from either 1s-3d quadrupole transitions or dipole-like transitions due to hybridized d-p states, depending on the site symmetry \cite{DeGroot20091s}. For Ni atoms in $\rm La_3Ni_2O_7$ with 4mm (C4v) point group symmetry, the pre-edge peak \textbf{A} can be attributed to dipole-allowed transitions from 1s to hybridized 3d/4p states. Indeed, comparison of XANES calculations with and without the quadrupole contribution (see inset of Fig. \ref{FDMNS}) shows that the pre-edge feature primarily originates from dipole-type transitions. Final states correspond to these transitions are empty or partially filled bound states below the continuum. Consequently, pre-edge peaks typically do not shift under external pressure \cite{Chen2019revisiting, Mijit2021crystal}. If there was a change in the oxidation state of Ni, a distinct chemical shift in peak \textbf{A} would be expected [as shown in Fig. \ref{Ref}(a)]. Thus, the stable energy position of pre-edge peak \textbf{A} provides another solid evidence that the average Ni valence in $\rm La_3Ni_2O_7$ remains constant throughout the investigated P-T range. Since the possibility of changes in oxidation and spin states are already ruled out, the slight increase in the intensity of \textbf{A} can be attributed to possible increase of orbital hybridization under high pressure. Note also that small increase (or fluctuation) of the pre-edge intensity was observed also in other 3d transition metal oxides under pressure, without variations in the valence and spin states \cite{Chen2019revisiting, Mijit2021crystal}. 

As shown in Fig. \ref{FDMNS}, feature \textbf {C} was reproduced with a minimum of 77 atoms in the cluster, requiring the inclusion of four apical O atoms at a distance of R = 5.684 \AA. This suggests that the enhancement of peak \textbf{C} under pressure has a non-local origin and is associated with increased scattering amplitude (around $\rm E \sim 8358$ eV) from multiple scattering pathways involving these apical oxygen atoms. It is noteworthy that the pressure-induced spectral changes in Fig. \ref{Fig_2}, particularly the enhancement of feature \textbf{C}, closely resemble the modifications observed in the Ni \textit{K}-edge XANES across the insulator-metal transitions of certain $\rm RNiO_3$ compounds (R = Pr, Nd, Y) \cite{Medarde1992,Acosta2008,Ramos2012,Rodrigues2023mapping}. This similarity maybe related with the increase of Ni-O-Ni bond angle along c-axis, a phenomenon common to both $\rm La_3Ni_2O_7$  \cite{Sun2023, Wang2024structure} and $\rm RNiO_3$ compounds \cite{Torrance1992,Wang2024structure}. This explanation is in agreement with the XANES simulations, which revealed a link between peak \textbf{C} and the apical oxygen atoms. The success of the spectral simulation suggests that the $\rm I4/mmm$ structure (refined from XRD measurements \cite{Wang2024structure}) reasonably captures the short and medium range structure of $\rm La_3Ni_2O_7$.

\begin{figure}[!]
\centering
\includegraphics[width=0.95\linewidth]{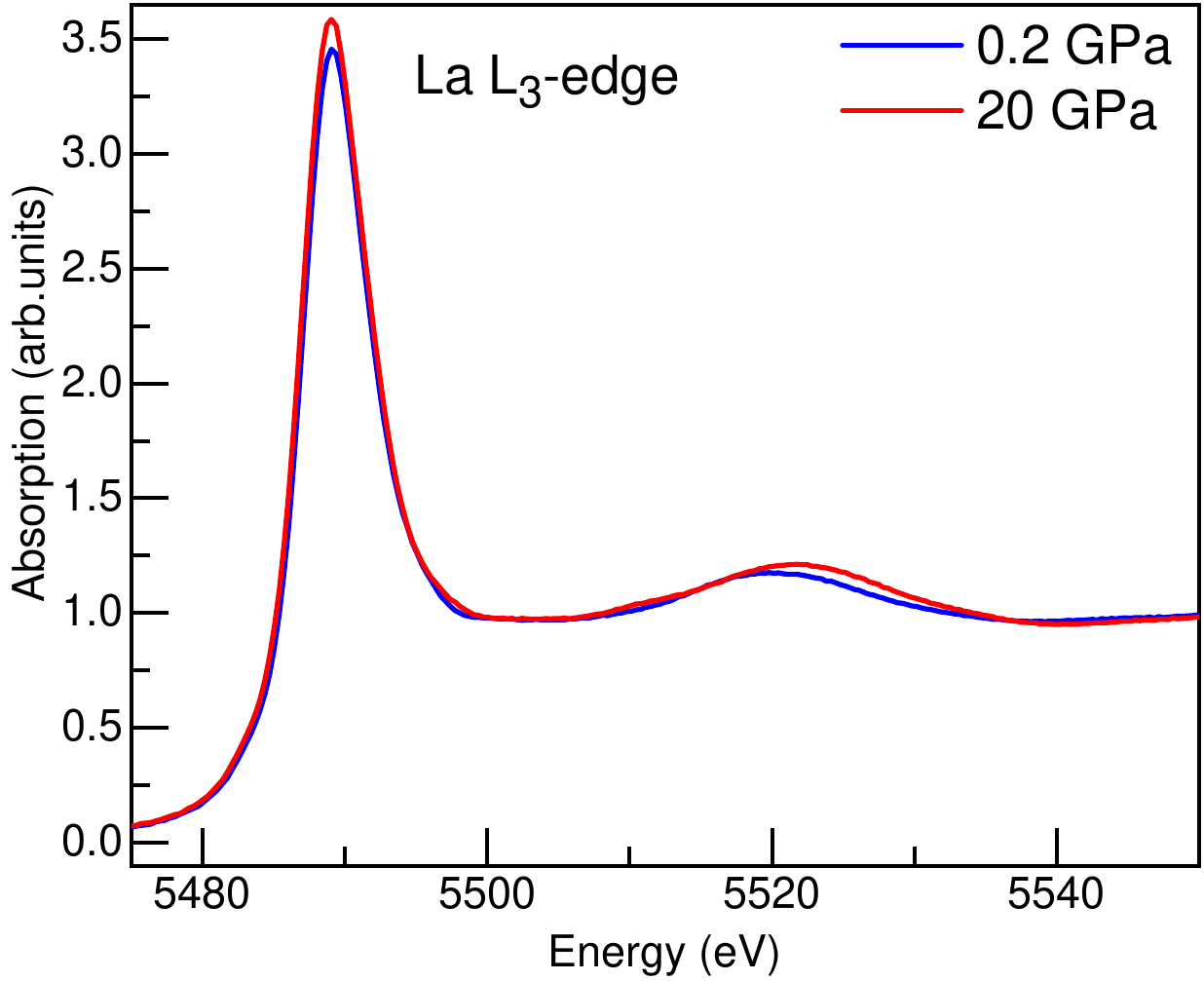}
\caption{La $L_3$-edge XANES.} 
\label{La_L3}
\end{figure}

La \textit{$L_3$}-edge XANES spectra recorded for ambient and high pressure phases are compared in Fig. \ref{La_L3}. Data measured at intermediate pressures are included in Fig. S3 of the SI file. As shown in Fig. \ref{La_L3}, the changes in the La \textit{$L_3$}-edge XANES are relatively subtle. Contrary to Ni K-edge XAS, a slight negative shift (-0.22 eV) is observed at the rising edge. However, the broad resonance near ~5520 eV shifts to higher energy following the typical behavior (compression effect). The negative shift at the rising edge clearly indicates a pressure-induced lowering of the unoccupied 5d levels. Additionally, we observed a modest increase ($\sim 4\%$) in the intensity of the white line peak under pressure, suggesting an enhancement of the sharp La 5d density of states just above the Fermi level.

\section{Conclusion}
The X-ray spectroscopy studies presented here provide valuable insights into the local electronic properties of \(\rm La_3Ni_2O_7\) in both normal and superconducting states.
Ni \textit{K}-edge XAS investigations reveal that the average oxidation state of Ni ions remains unaffected under varying pressure (up to 25 GPa) and temperature (down to 10 K) conditions, staying fixed at the nominal value of approximately \(2.5+\). Ni \(K_{\beta}\) XES provides compelling evidence for a low-spin (LS, \(\rm S=1/2\)) ground state with strong charge transfer character in \(\rm La_3Ni_2O_7\). High-pressure XES measurements confirm that the LS ground state is stable under pressure, effectively ruling out previously proposed spin transition scenarios \cite{Jiang2024pressure,Labollita2023electronic,Liao2023electron}. 
The pressure response of the valence-to-core \(K_{\beta_{2,5}}\) emission indicates no significant changes in the oxygen \(2p\)-character mixed with Ni \(3d\) states under compression. The success of our XANES simulations suggests that the $\rm I4/mmm$ structure (obtained from XRD measurements \cite{Wang2024structure}) reasonably captures the local structure of \(\rm La_3Ni_2O_7\). In contrast to the Ni \textit{K}-edge, the La \textit{$L_3$}-edge XANES exhibits only subtle pressure-induced changes, indicating minimal modifications at the local La sites. A slight negative shift (-0.22 eV) of the \textit{$L_3$}-edge suggests a pressure-induced lowering of the 5d density of states above the Fermi level. The implications of this for superconductivity require further theoretical investigations.

During the preparation of this manuscript, we became aware of another high-pressure XAS study on the same compound, which was recently posted on arXiv \cite{Li2024Dist}. Contrary to the findings in this paper, we did not observe noticeable changes in the oxidation and spin state in $\rm LaNiO_3$ within the investigated P-T range. 

\section {Experiments and Methods}
\subsection {Sample preparation:} \ $\rm La_3Ni_2O_7$ samples investigated in this work were provided by the authors of ref. \cite{Sun2023} in the powder form. We refer refs. \cite{Sun2023,wang2024bulk} for the details of sample synthesis. Phase purity of the sample was confirmed by x-ray powder diffraction measurement under ambient condition (performed at the ID15b beamline of ESRF \cite{Garbarino2024}), see Fig. S1 in the SI file. 

\subsection {XAS and XES measurements:} \  XAS measurements at Ni \textit{K} and La $L_3$ edges have been collected at the BM23 and ID24 beamlines of ESRF\cite{Mathon2015,Rosa2024}, along five independent runs. At both XAS beamlines (BM23 and ID24), x-ray beam was  monochromatized with a Si(111) double crystal monochromator and focused down to 3 × 3 $\rm \mu m^2$ (FWHM) using two Pt-coated mirrors in the standard Kirkpatrick–Baez geometry that serves also the rejection of higher harmonics. Ni $K_{\beta}$ XES measurements (run-6) have been performed at the ID20 beamline of ESRF \cite{Sahle2024id20,Moretti2018high}. Incident x-ray beam at 10 keV was monochromatized using Si(111) monochromator and focused by a Kirkpatrick-Baez mirror to 10 × 10 $\rm \mu m^2$ (FWHM) spot size. Emission signal was collected at near-$90\degree$ scattering angle with an energy-dispersive von Hamos-type spectrometer \cite{sahle2023compact}. 

\subsection{Sample environment:}
\ High pressure was generated using a membrane-driven diamond anvil cells (DACs). DACs used for XAS measurements were equipped with nanopolycrystalline diamond (NPD) anvils \cite {Irif2003,Ishi2012} to obtain glitch free high quality XAS data. Low temperatures down to 10 K was achieved using a He flow cryostat. Perforated (both sides) NPD anvils with 300 $\rm \mu m$ culet were used for the first and third run. While NPD anvils (one side is perforated) of 600 $\mu m$ culet were used for the second experimental run. We used silicon oil as pressure transmitting medium (PTM) for these measurements. Forth and fifth compression run at Ni $K$ and La $L_3$-edges were performed under room temperature at the ID24 XAS beamline \cite{Rosa2024}, taking advantage of the high brilliance from the undulator source. Perforated (both sides) NPD anvils with 300 $\rm \mu m$ culet was used for these runs and Daphne 7575 oil was used a PTM. Note that experiments with NPD anvils are not compatible with optimal PTMs like He gas, as He easily penetrates into the microstructures of the NPD anvil. Sample loading with other gases (like Neon) would complicate the experiments at low temperature. Therefore, liquid PTMs (Si or Daphne oils) were considered as a convenient choice, also to better reproduce the experimental conditions of ref. \cite{Sun2023} which were achieved mostly under quasi-hydrostatic conditions using solid PTMs. For the XES measurements under pressure (run-6), sample was loaded in a panoramic DAC equipped with single crystal anvils of  500 $\mu m$, without any pressure transmitting medium. 

\subsection {XANES simulations: } \ theoretical Ni \textit{K}-edge XANES (x-ray absorption near edge structure) spectra were simulated using the real-space FDMNES code \cite{FDMNS,Joly2001, Joly2009}. We used the multiple scattering theory mode of the code. Calculations were self-consistent and relativistic (neglecting the spin-orbit) in order to have a better electronic structure probed by the photoelectron, as well as a better determination of the Fermi Level. In XAS only states above it are probed. Both dipole and quadrupole transitions were considered. At the Ni K-edge, they probe respectively the Ni-4p and Ni-3d states. The broadening to take into-account the time-life of the core and photoelectron states are done, in a standard way, by a Lorentzian convolution, with an energy dependent width.

\section{Acknowledgements}
\begin{acknowledgments}
This study was supported financially by National Key Research and Development Program of China (Grant No. 2023YFA1406000), the National Natural Science Foundation of China (Grant Nos. 22171283 and 22203031). We acknowledge the European Synchrotron Radiation Facility for provision of synchrotron radiation facilities (proposal No: HC-5754 and HC5755). We would like to warmly thank J. Jacobs for his help with the preparation of the diamond anvil cells. Work at SYSU was supported by the National Natural Science Foundation of China (Grants No. 12425404 and 12174454), the National Key Research and Development Program of China (Grant No. 2023YFA1406500), the Guangdong Basic and Applied Basic Research Funds (Grant No. 2024B1515020040), Guangzhou Basic and Applied Basic Research Funds (Grant No. 2024A04J6417), and Guangdong Provincial Key Laboratory of Magnetoelectric Physics and Devices (Grant No. 2022B1212010008). 
\end{acknowledgments}

\nocite{*}

\bibliography{ref_v2}

\end{document}


\begin{center} 
{\LARGE \bf{Supplementary Information\\}}
\end{center}

\vspace{2 cm}
\begin{center} 
{\Large \bf{Local electronic properties of $\rm La_3Ni_2O_7$ under pressure}}
\end{center}

\vspace{2 cm}


\begin{figure*}[h]
\centering
\includegraphics[width=0.7\linewidth]{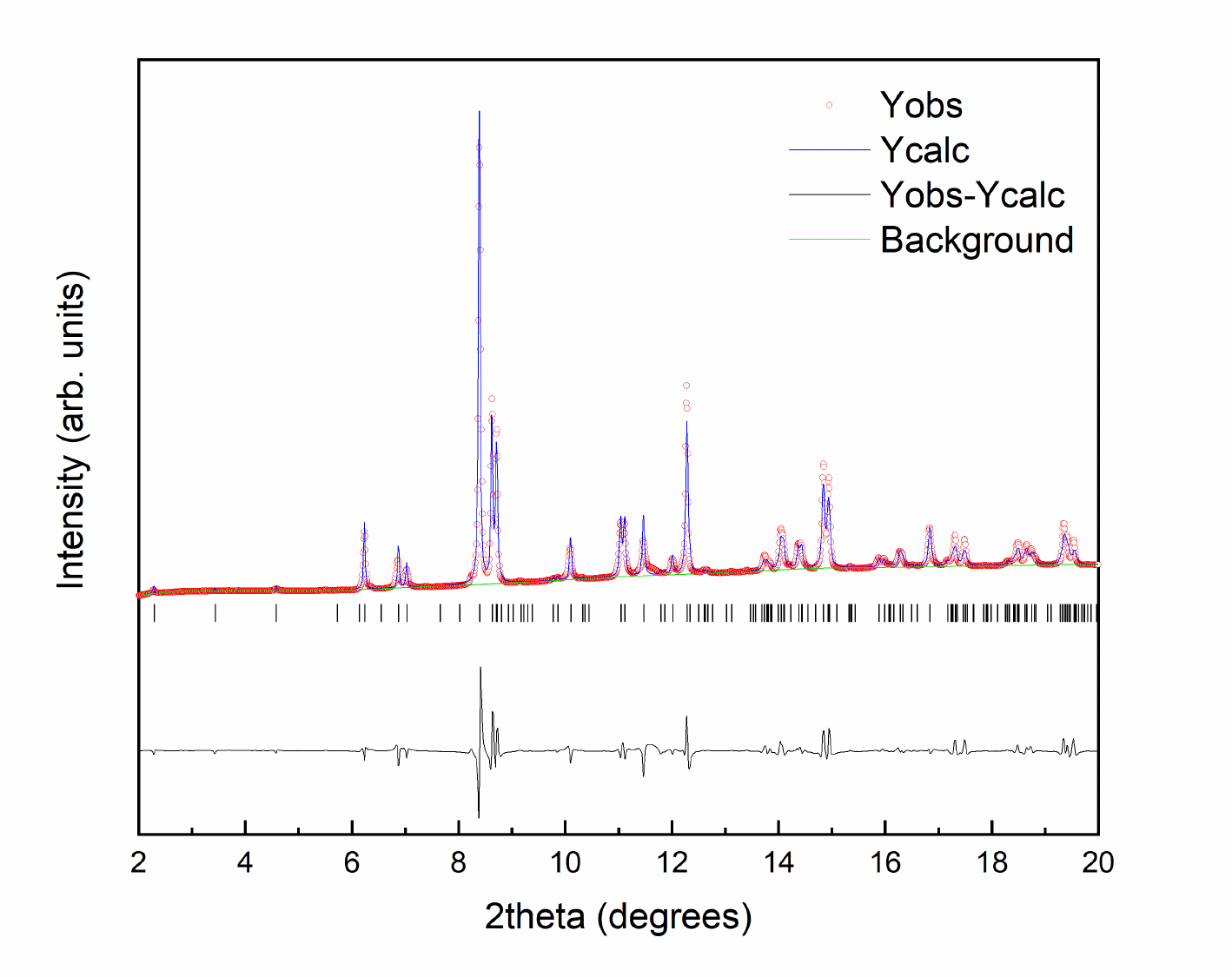}
\caption{Powder X-ray diffraction measurements at ambient conditions confirm that the sample is phase-pure $\rm La_3Ni_2O_7$. The diffraction pattern was fitted using the ambient pressure orthorhombic structure with the Cmmm space group. The refined lattice parameters (a = 5.38902 Å, b = 5.44791 Å, c = 20.50644 Å) are in good agreement with previously reported literature values. Diffraction pattern was collected at ID15b beamline of ESRF with x-ray photon energy at 30 keV ($\lambda \sim 0.41 \ \AA$).}
\label{Fig_S1}
\end{figure*}

\begin{figure}[!]
\centering
\includegraphics[width=0.6\linewidth]{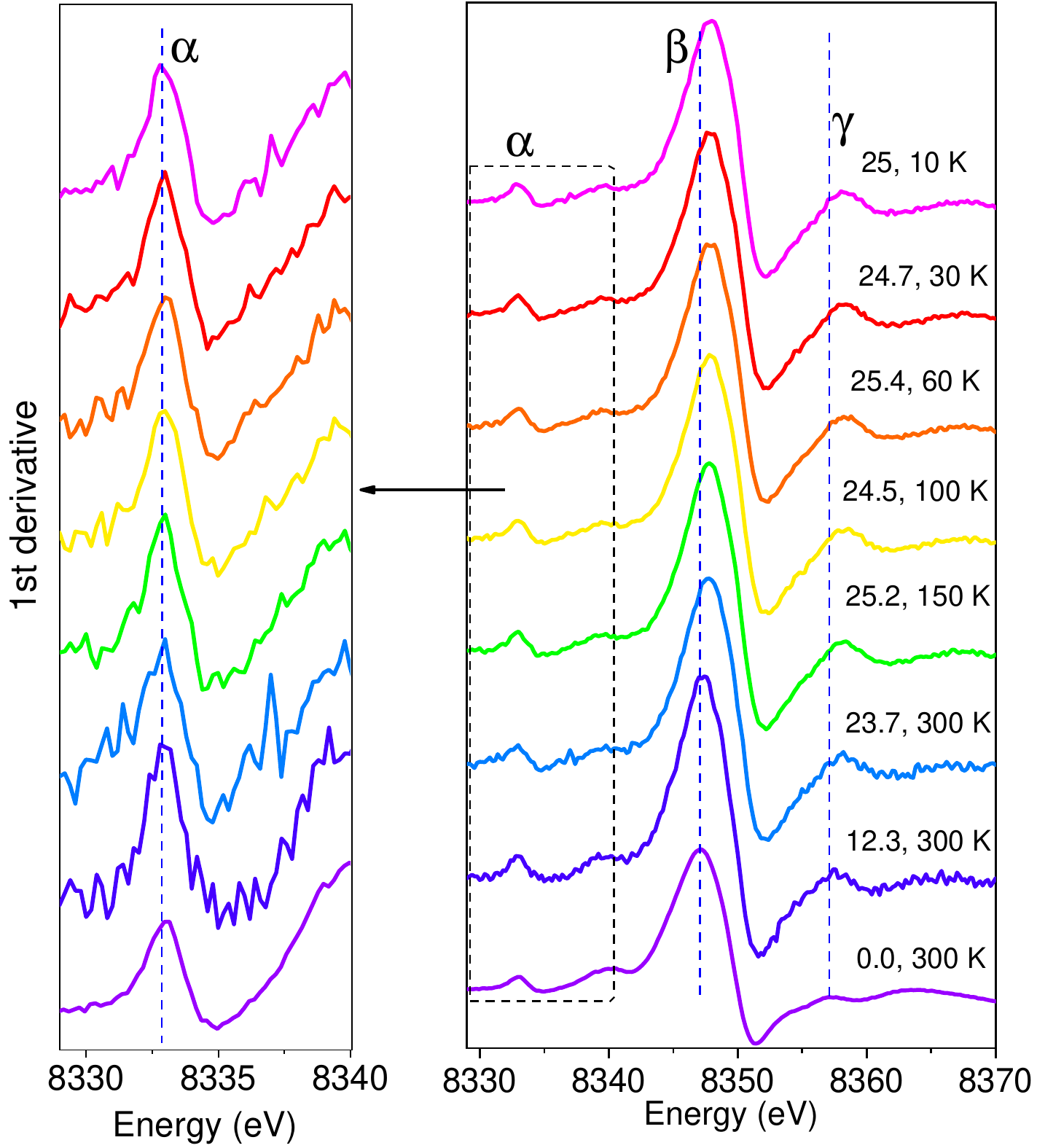}
\caption{First derivative of the Ni \textit{K}-edge XANES measured along the 2nd experimental run (run-2). Pre-edge range was shown with magnification at the left panel. Stable energy position and nearly similar line shapes of pre-edge peak \textbf{A} can be seen from the behavior of feature \textbf{$\alpha$} in the first derivative spectra (left panel). Shift of the edge energy is evident from the movement of the main peak (feature \textbf{$\beta$}), while the enhancement of feature \textbf{C} is captured by the appearance of feature \textbf{$\gamma$} in the first derivative spectra (right panel). The absence of spectral modifications upon cooling at $23.7 \sim 25$ GPa (see Fig. 4d in the main text) is further confirmed by the identical line shapes of the first derivative spectra.} 
\label{Run3}
\end{figure}

\begin{figure}[!]
\centering
\includegraphics[width=0.4\linewidth]{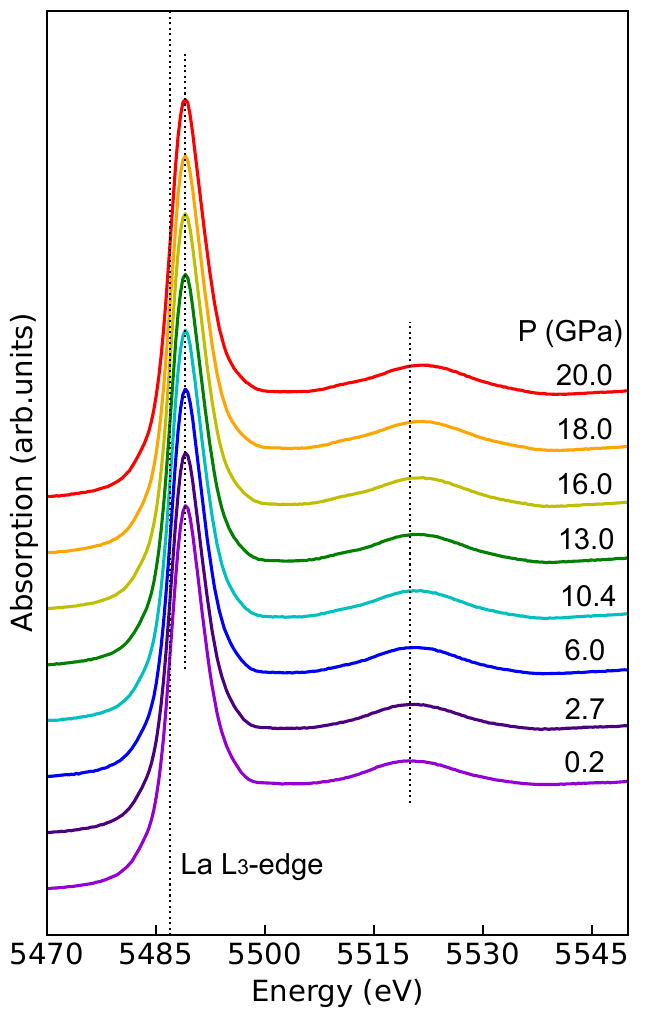}
\caption{La $L_3$-edge XANES of $\rm La_3Ni_2O_7$ measured along run-5 up to 20 GPa under room temperature. } 
\label{La_L3}
\end{figure}
